\documentclass{article}
\usepackage[a4paper, left=1.5cm, right=1.5cm, top=2.5cm, bottom=2.5cm]{geometry}
\usepackage{graphicx} 
\usepackage{amsmath}
\usepackage{amssymb}
\usepackage{xcolor}
\usepackage{authblk} 
\usepackage{cite}

\title{Could Planck Star Remnants be Dark Matter?}

\author[1,2]{Oem Trivedi\thanks{oem.trivedi@vanderbilt.edu}}
\author[3]{Abraham Loeb\thanks{aloeb@cfa.harvard.edu}}

\affil[1]{Department of Physics and Astronomy, Vanderbilt University, Nashville, TN 37235, USA}
\affil[2]{International Centre for Space and Cosmology, Ahmedabad University, Ahmedabad 380009, India}
\affil[3]{Astronomy Department, Harvard University, 60 Garden St., Cambridge, MA 02138, USA}

\date{\today}

\begin{document}

\maketitle

\begin{abstract}
We explore the end state of gravitational collapse under quantum gravity effects and propose that Planck Star Remnants (PSR), formed via nonsingular bounces, could serve as viable dark matter candidates. Within the framework of Loop Quantum Cosmology, we model the collapse of a homogeneous matter distribution and show that the classical singularity is replaced by a quantum bounce at the Planck density. By analytically matching the Friedman-Lemaître–Robertson–Walker (FLRW) interior to an exterior Schwarzschild spacetime using the Israel junction conditions, we demonstrate that the bounce remains causally hidden from external observers, avoiding any observable re-expansion. This naturally leads to the formation of stable, non-radiating PSR, whose radius coincides with the Schwarzschild radius when the black hole mass approaches the Planck mass as a result of Hawking evaporation. We suggest that such remnants may originate from evaporating primordial black holes in the early universe, and estimate the relic abundance needed for PSR to account for the observed dark matter density. We also discuss some crucial differences between PSR and previous proposals of Planck mass relics. The scenario is shown to be consistent with existing astrophysical and cosmological constraints, offering a unified framework connecting quantum gravitational collapse, and the nature of dark matter.
\end{abstract}

\section{Introduction}

One of the most profound unresolved questions in gravitational physics is the fate of matter undergoing complete gravitational collapse. In classical general relativity, such collapse inevitably leads to a singularity \cite{sb1hawking1972black,sb2bekenstein1973black,sb3bekenstein1974generalized,sb4tipler1977singularities,sb5krolak1978singularities,sb6clarke1985conditions,sb7penrose1969gravitational}. It is widely expected that quantum gravitational effects become important at the Planck density of $(c^5/h G^2)\sim 10^{93}~{\rm g~cm^{-3}}$ and could resolve the singularity through new dynamics \cite{sb8joshi2002cosmic,sb9senovilla20151965,sb10penrose1965gravitational,sb11penrose1999question,sb12hawking1970singularities,sb13Ong:2020xwv,sb14Vagnozzi:2022moj,sb15joshi2011recent,sb16janis1968reality,sb17Joshi:2011zm}. Loop Quantum Gravity (LQG) \cite{lqg3fRovelli:1994ge,lqg1Ashtekar:2004eh,lqg2Garay:1994en,lqg4Ashtekar:2011ni,lqg5Bern:2010ue,lqg6Rovelli:1997yv}, and in particular its application to cosmology in the form of Loop Quantum Cosmology (LQC) \cite{lqc1Ashtekar:2006rx,lqc2Agullo:2016tjh,lqc3Bojowald:2005epg,lqc4Ashtekar:2011ni,lqc5Agullo:2023rqq,lqc6Ashtekar:2008zu,lqc7Banerjee:2011qu,lqc8Wilson-Ewing:2016yan,Nojiri:2005sx}, has provided an explicit manifestation of this expectation, replacing singularities with non-singular bounces for both cosmological and black hole contexts.
\\
\\
While LQC-inspired bounce models offer a compelling way to avoid singularities, a central question is whether such bounces can have observable consequences. Specifically, if a black hole interior undergoes a bounce at high curvature and transitions into an expanding geometry, can this lead to an observable white hole or some form of explosive outflow? In this work, we analyze this scenario in detail using both the effective LQC dynamics for a collapsing homogeneous interior and a rigorous treatment of the matching conditions to an exterior Schwarzschild geometry via the Israel junction conditions \cite{Israel:1966rt,Israel:1967zz}. We demonstrate that under reasonable assumptions, the bounce remains causally disconnected from an external observer and does not lead to any visible signature on astrophysical timescales. This may seem discouraging from an observational standpoint but it ends up being physically relevant. If matter collapses to the Planck density, undergoes a quantum bounce and yet remains hidden from external observers indefinitely, the most natural outcome is a stable, non-singular remnant. The existence of a remnant, labeled a Planck star, was first proposed in the context of LQG \cite{pl1Rovelli:2014cta,pl2Rovelli:2024sjl,pl3Rovelli:2017zoa,pl4Scardigli:2022jtt,pl5Wilson-Ewing:2024uad,pl6Barrau:2014hda,pl7Tarrant:2019tgv} as a replacement to the classical singularity with a quantum gravitational core whose radius is approximately equal to its Schwarzschild radius when the black hole mass equals the Planck mass, $\sqrt{h c/G}\sim 10^{-5}~{\rm g}$. Such remnants would behave externally like black holes, but would possess a finite size and non-singular interior protected by quantum gravity effects.
\\
\\
This leads to an intriguing cosmological application where Primordial black holes (PBHs),potentially formed in the early universe through large density fluctuations \cite{pbh1pbhzel1966hypothesis,pbh2hawking1971gravitationally,pbh3carr1974black,pbh4carr1975primordial,pbh5chapline1975cosmological,pbh6hawking1975particle,pbh7hawking1974black,pbh8khlopov1980primordial,pbh9polnarev1985cosmology,pbh10khlopov2010primordial,pbh11carr2016primordial,pbh12carr2020primordial,pbh13carr2024observational}, would have undergone Hawking evaporation over cosmic timescales. However, if evaporation halts at the Planck scale due to quantum gravitational backreaction, then such black holes would not vanish entirely but would leave behind Planck-mass relics. If formed in sufficient abundance, these relics could account for the cold dark matter observed today as the relics would be stable, compact and interact only gravitationally, matching the empirical requirements on the nature of dark matter particles.
\\
\\
This work is structured as follows. We begin by analyzing the dynamics of a collapsing homogeneous matter distribution in LQC, derive the matching conditions at the bounce and discuss causality aspects in \S 2. We then interpret the consequences of this inference in terms of Planck Star Remnants (PSR) and evaluate their viability as dark matter in \S 3, also considering the relic abundance of PSR and existing cosmological constraints. Finally, we summarize our main conclusions in \S 4.

\section{Quantum-Corrected Collapse}
To model the interior of a collapsing matter distribution near the Planck density, we adopt a closed Friedman-Lemaître–Robertson–Walker (FLRW) metric, modified by effective LQC corrections. This allows for singularity avoidance via a bounce, and serves as a tractable approximation of spherically-symmetric quantum-gravitational collapse. The spacetime inside the collapsing region is described by the line element (adopting the convention $c=1$),
\begin{equation}
    ds^2 = -dt^2 + a(t)^2 \left[ \frac{dr^2}{1 - kr^2} + r^2 d\Omega^2 \right] .
\end{equation}
Adopting a flat space approximation to find the quantum-corrected dynamics, the scale factor $a(t)$ is governed in LQC by the modified Friedmann equation \cite{lqc9MenaMarugan:2011va,lqc10MenaMarugan:2009ru},
\begin{equation}
    H^2 = \left( \frac{\dot{a}}{a} \right)^2 = \frac{8\pi G}{3} \rho \left(1 - \frac{\rho}{\rho_c} \right),
\end{equation}
where $\rho$ is the energy (mass) density and $\rho_c$ is the critical density set by the underlying loop quantum geometry, typically taken to be of order the Planck density. This equation reduces to the classical Friedmann equation at low densities but leads to a bounce when $\rho = \rho_c$ due to the vanishing of the Hubble rate, $H=\dot{a}/a$. We assume the matter to be pressureless ($p = 0$) \footnote{The pressure term in the continuity equation does not affect the forthcoming analysis on the bounce scenario. The pressure will be discussed in the context of the dissipative effects considered later.}, consistent with a Planck density phase dominated by kinetic energy, so that,
\begin{equation}
    \dot{\rho} + 3H\rho = 0,
\end{equation}
which yields $\rho(a) \propto a^{-3}$. The interior metric is matched at the boundary $r = r_0$ to an exterior Schwarzschild geometry for a total enclosed mass $M$,
\begin{equation}
    ds^2 = -\left(1 - \frac{2GM}{R}\right) dT^2 + \left(1 - \frac{2GM}{R}\right)^{-1} dR^2 + R^2 d\Omega^2.
\end{equation}
The Israel junction matching conditions require the continuity of the induced metric and the extrinsic curvature across the boundary hypersurface \cite{Israel:1966rt,Israel:1967zz} $\Sigma$. The induced metric continuity yields
\begin{equation}
    R(t) = r_0 a(t),
\end{equation}
identifying the Schwarzschild areal radius $R$ at the boundary with the proper radius of the collapsing sphere. We also note that while $r_0$ is indeed constant and labels the comoving boundary of the FLRW region, the actual matching is done at a constant-time hypersurface across the bounce. Our setup assumes that quantum gravitational effects are significant only in the high-density interior region \cite{lqg7Rovelli:1989za,lqg8BarberoG:1994eia,lqg9Amelino-Camelia:2008aez} and become negligible outside the collapsing matter distribution. In particular, the Schwarzschild exterior solution remains a valid classical vacuum solution of General Relativity, since the quantum modifications in LQC are tied to regions of high curvature or Planck energy density. In this context, the vacuum exterior has a sub-Planckian curvature with a vanishing energy-momentum tensor and is not expected to receive significant corrections from quantum gravity effects. As such, matching a quantum-corrected FLRW interior to a classical Schwarzschild exterior \footnote{While our analysis assumes a non-rotating, spherically sym.metric collapse,which is mostly fine for PBHs we note that extending this framework to include angular momentum remains a major challenge. The matching of a rotating interior to an exterior Kerr or cosmologically modified Kerr geometry is a nontrivial problem \cite{kds1Agnese:1999df,kds2Akcay:2010vt}.} is physically justified within the effective field theory approach \footnote{For details on how Effective Field Theory methods are employed, see \cite{eft1Penco:2020kvy,eft2Burgess:2007pt,eft3Manohar:2018aog}.}, where quantum gravity effects are localized and do not propagate beyond the matter distribution.  
\\
\\
To impose the Israel junction conditions, we compute the extrinsic curvature $K_{ij}$ on both sides of the boundary. In the FLRW interior, the relevant non-zero components of the extrinsic curvature tensor evaluated at $r = r_0$ are
\begin{align}
    K_{\theta\theta}^{\text{FLRW}} &= -r_0^2 a(t)^3 \dot{a}(t) ,\\
    K_{\phi\phi}^{\text{FLRW}} &= -r_0^2 a(t)^3 \dot{a}(t) \sin^2\theta ,\\
    K_{rr}^{\text{FLRW}} &= -a(t)\dot{a}(t).
\end{align}
For the Schwarzschild exterior, the non-zero extrinsic curvature components for a boundary surface at $R(t)$, moving in the exterior coordinates, are more involved but reduce to expressions involving $M$, $R(t)$, and $\dot{R}(t) = r_0 \dot{a}(t)$
\begin{align}
    K_{rr}^{\text{Sch}} &= -\frac{M}{r_0^2 a^2 (1 - \frac{2M}{r_0 a})} + \frac{2M}{r_0^2 a^2 (1 - \frac{2M}{r_0 a})^2}, \\
    K_{\theta\theta}^{\text{Sch}} &= -r_0 a(t) \left(1 - \frac{2M}{r_0 a(t)} \right) + 2 r_0 a(t), \\
    K_{\phi\phi}^{\text{Sch}} &= \sin^2\theta \cdot K_{\theta\theta}^{\text{Sch}}.
\end{align}
We would also like to clarify about the nature of extrinsic curvatures here. Because our matching surface is spacelike (at fixed proper time in the interior), the relevant components of the extrinsic curvature are the spatial components (such as $K_{rr}, K_{\theta\theta}$, etc.). The appearance of $K_{rr}$ is thus natural and necessary in our context and one would be right to think that if we were matching across a timelike worldtube then $K_{tt}$ and angular components would be the primary ones involved but here we are matching across a constant-time spatial slice, and hence the inclusion of $K_{rr}$ and the like is appropriate \footnote{For more details on these conditions, one can see \cite{Grenon:2009sx,Bambi:2013caa}.}. The junction condition $[K_{ij}] = 0$ implies that the difference in these quantities across $\Sigma$ must vanish and this leads to two coupled dynamical equations:
\begin{align}
    -\frac{M}{r_0^2 a^2 (1 - \frac{2M}{r_0 a})} + \frac{2M}{r_0^2 a^2 (1 - \frac{2M}{r_0 a})^2} + a \dot{a} &= 0 \label{eq:radial-junction}, \\
    -r_0 a \left(1 - \frac{2M}{r_0 a} \right) + 2 r_0 a + r_0^2 a^3 \dot{a} &= 0 .\label{eq:tangent-junction}
\end{align}
These equations ensure a smooth geometric transition from the interior quantum-corrected FLRW region to the classical Schwarzschild exterior. This leads to the following system of coupled differential equations:
\begin{equation}
\begin{cases}
\dot{a} = a \sqrt{\frac{8\pi G}{3} \rho \left(1 - \frac{\rho}{\rho_c} \right)}, \\
-\frac{M}{r_0^2 a^2 \left(1 - \frac{2M}{r_0 a}\right)} + \frac{2M}{r_0^2 a^2 \left(1 - \frac{2M}{r_0 a}\right)^2} + a \dot{a} = 0, \\
-r_0 a \left(1 - \frac{2M}{r_0 a} \right) + 2r_0 a + r_0^2 a^3 \dot{a} = 0.
\end{cases}
\end{equation}
By solving this system numerically, we determine the evolution of $R(t)$ and confirm whether the LQC bounce prevents $R(t) \to 0$. 
\begin{figure}[!h]
    \centering
    \includegraphics[width=1\linewidth]{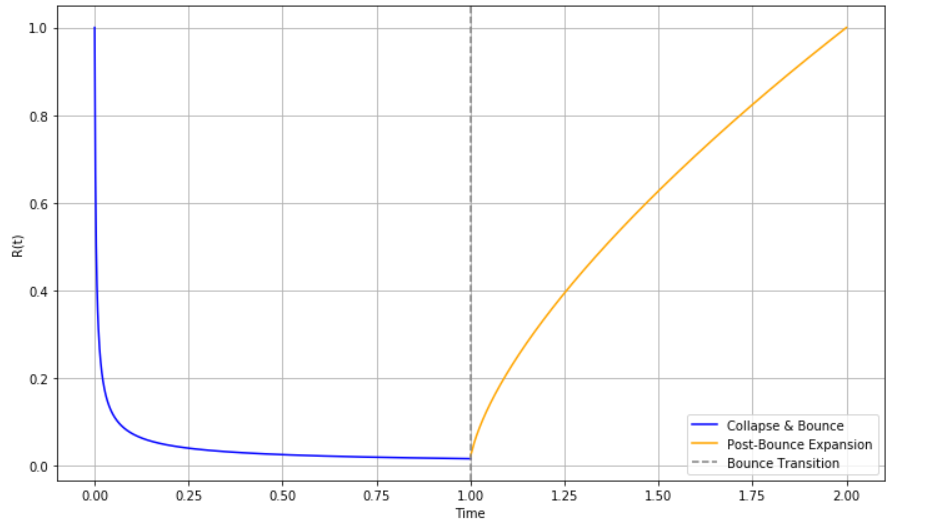}
    \caption{Normalized evolution of the the boundary radius of the Schwarzschild exterior during the contraction and the post bounce phase of the interior matter distribution.}
    \label{lqc}
\end{figure}
Figure \ref{lqc} shows the evolution of the collapsing sphere during the contraction phase governed by the FLRW metric under LQC \footnote{The radius is normalized by its initial value $R(t=0)$, and time is expressed in Planck units $t / t_{\text{Pl}}$ so this normalization allows for dimensionless analysis of the collapse and bounce dynamics and allows one to scale both epochs properly in the figure.}. The radius of the collapsing sphere never reaches zero and the sphere begins to expand again after the bounce. This implies that the Schwarzschild exterior never collapses to a central singularity.
\\
\\
For the sake of completeness, we also explore whether bulk dissipative effects can modify the matching conditions discussed above by considering corrections to the interior FLRW dynamics from a phenomenological dissipation term, modeled as bulk viscosity. In such scenarios, the effective pressure is modified as \cite{Hu:2020xus}
\begin{equation}
    P_{\text{eff}} = p - 3 \zeta H,
\end{equation}
where $ \zeta $ is the bulk viscosity coefficient and the modified continuity equation in the presence of dissipation becomes
\begin{equation}
    \dot{\rho} + 3H(\rho + p - 3 \zeta H) = 0.
\end{equation}
The Raychaudhuri equation, after incorporating LQC corrections and dissipation, takes the form \cite{lqc1Ashtekar:2006rx,lqc2Agullo:2016tjh,lqc3Bojowald:2005epg,lqc4Ashtekar:2011ni,lqc6Ashtekar:2008zu}
\begin{equation}
    \dot{H} = -4 \pi G ( \rho + p - 3 \zeta H ) \left( 1 - \frac{2 \rho}{\rho_c} \right).
\end{equation}
To gauge how dissipation alters the collapse, we expand the Hubble parameter and the energy density perturbatively in the dissipation parameter $ \zeta $,
\begin{align}
    H(t) &= H_0(t) + \zeta  \delta H(t) + \mathcal{O}(\zeta^2), \\
    \rho(t) &= \rho_0(t) + \zeta  \delta \rho(t) + \mathcal{O}(\zeta^2),
\end{align}
where $ H_0(t) $ and $ \rho_0(t) $ are the zeroth-order (no-dissipation) solutions governed by
\begin{align}
    H_0^2 &= \frac{8\pi G}{3} \rho_0 \left( 1 - \frac{\rho_0}{\rho_c} \right), \\
    \dot{\rho}_0 &= -3H_0 (\rho_0 + p), \\
    \dot{H}_0 &= -4 \pi G (\rho_0 + p) \left(1 - \frac{2 \rho_0}{\rho_c} .\right)
\end{align}
Substituting the perturbative expansion into the dissipative Raychaudhuri equation and retaining only first-order terms, we obtain
\begin{equation}
    \dot{H} \approx \dot{H}_0 + \zeta \left( \frac{\partial \dot{H}}{\partial \zeta} \right)_{\zeta = 0},
\end{equation}
with the first-order correction given by
\begin{equation}
    \frac{\partial \dot{H}}{\partial \zeta} \Big|_{\zeta = 0} = 12 \pi G H_0 \left(1 - \frac{2 \rho_0}{\rho_c} \right),
\end{equation}
leading to
\begin{equation}
    \dot{H} \approx \dot{H}_0 + 12 \pi G \zeta H_0 \left(1 - \frac{2 \rho_0}{\rho_c} \right).
\end{equation}
Similarly, 
\begin{equation}
    \dot{\rho} = \dot{\rho}_0 + \zeta \left( \frac{\partial \dot{\rho}}{\partial \zeta} \right)_{\zeta = 0},
\end{equation}
with
\begin{equation}
    \frac{\partial \dot{\rho}}{\partial \zeta} \Big|_{\zeta = 0} = 9 H_0^2,
\end{equation}
yielding the correction,
\begin{equation}
    \dot{\rho} \approx \dot{\rho}_0 + 9 \zeta H_0^2.
\end{equation}
The first-order correction to the Hubble parameter can then be expressed formally as
\begin{equation}
    \delta H(t) = 3 \pi G \int H_0(t') \left(1 - \frac{2 \rho_0(t')}{\rho_c} \right)  dt',
\end{equation}
and the correction to the energy density becomes
\begin{equation}
    \delta \rho(t) = 9 \int H_0(t')^2  dt'.
\end{equation}
Hence, dissipation slightly reduces the contraction rate and slows the increase of energy density, acting as a mild drag term. However, as $ \rho_0 \to \rho_c $ the integrands are suppressed by the LQC correction factors, and dissipation becomes negligible. Hence, unless $ \zeta $ is unphysically large, these corrections do not prevent the bounce or significantly alter the trajectory.\footnote{If the dissipated energy escapes the system as gravitational radiation then it must be modeled as a separate energy flux through the junction surface. In our model, we have ignored energy flux through the boundary.} 
\\
\\
From the perspective of the interior FLRW observer, the bounce and re-expansion happen rapidly. However, to an external Schwarzschild observer the event horizon imposes a causal barrier. This means that the re-expansion remains hidden unless spacetime geometry is globally altered and thus for matter to emerge visibly, one must invoke a new asymptotic region \footnote{An example of such a region would be that of a white hole \cite{wh1Bianchi:2018mml,wh2Han:2023wxg,wh3Nikitin:2018omt,wh4Gaur:2023ved}.} or permit energy to tunnel through a quantum-modified causal boundary. Without such mechanisms, the post-bounce phase is observationally invisible. While quantum gravitational corrections like those from LQC may prevent singularity formation internally, the exterior observability of the bounce depends crucially on how matter is allowed to re-emerge, and whether the causal structure permits it. 
\\
\\
It is important to address a subtle aspect of the matching procedure concerning the behavior of the Schwarzschild metric inside the horizon. For $r < 2GM$, the standard Schwarzschild coordinates exhibit a signature flip wherein the radial coordinate becomes timelike and the time coordinate becomes spacelike, which leads to an apparent mismatch with the usual FLRW coordinate structure. However this does not pose an obstruction to our matching analysis as the Israel junction conditions \cite{Israel:1966rt,Israel:1967zz} are formulated in terms of the induced 3-metric and the extrinsic curvature on a chosen hypersurface, which in our case is taken to be a constant-time spacelike slice across the bounce. These quantities are geometric and coordinate-invariant, and their continuity (or controlled discontinuity) determines the validity of the matching. This means that the signature change in Schwarzschild coordinates is a coordinate artifact and does not obstruct a valid physical matching and this point has been reinforced by in recent studies such as \cite{bounceexclusion}, where the authors demonstrate a successful matching of a collapsing FLRW cloud to an exterior Schwarzschild solution even deep inside the horizon using Lemaître-type coordinate systems. Therefore, we can say that our procedure of matching a quantum-corrected FLRW interior to a Schwarzschild exterior remains valid throughout the evolution including when the system is entirely within its gravitational radius.
\\
\\
We would also like to emphasize that the classical picture of inevitable collapse inside the Schwarzschild radius does not constrain the interior evolution in our model. In the quantum-corrected FLRW framework motivated by loop quantum effects that we considered, the evolution proceeds with respect to a globally well-defined proper time parameter associated with comoving observers. This evolution remains timelike and regular across the bounce, with the scale factor $a(t)$ (and hence the areal radius $r(t) = r_0 a(t)$) reaching a nonzero minimum when the density approaches the critical scale $\rho_c$ rather than decreasing indefinitely toward a singularity. The quantum geometry effects modify the effective Friedmann and Raychaudhuri equations in a way that halts the contraction and triggers re-expansion in proper time even when $r(t) < 2GM$. This means that while the classical Schwarzschild solution implies that the radius must shrink monotonically in forward time inside the horizon, our setup demonstrates that this behavior is altered by quantum gravity and the interior does not obey the same causal structure. This supports the interpretation of the bounce as producing a Planck-scale static remnant whose internal dynamics are not governed by classical singularity theorems, and hence we remain consistent with the modified gravitational framework we are applying.
\section{Planck Star Remnants as Dark Matter Candidates}
In the bounce solution, Planck star remnants arise from non-singular quantum-corrected gravitational collapse. The Planck star scenario offers a compelling endpoint for the LQC-regulated dynamics studied earlier. The bounce halts the collapse at near-Planck densities, thus freezing the evolution and leaving behind a stable compact object. Once Hawking evaporation reduces the mass $M$ to the Planck mass \footnote{We note that semiclassical arguments based on black hole entropy suggest that deviations from standard Hawking evaporation may begin well before the Planck mass is reached, most notably around the Page time, when half the initial entropy has been radiated. One such proposed modification is the recently advocated memory burden effect \cite{mbDvali:2024hsb}, which could significantly alter the late time dynamics and potentially prevent the formation of stable remnants. While our analysis focuses on the end stage dynamics near the Planck scale where loop quantum gravity effects dominate, we acknowledge that incorporating earlier quantum corrections such as those suggested by the memory burden mechanism is an important open direction for future work.}, the bounce radius becomes of order the Schwarzschild radius and the long-lived relic could make dark matter candidates. One could ask as to why would Hawking evaporation stop and make a stable Planck-mass relic? In standard semiclassical theory Hawking evaporation proceeds down to arbitrarily small scales but in quantum gravity scenarios like ours, this behavior is expected to be modified. In this case what happens is that as the black hole approaches the Planck mass, quantum gravitational pressure can halt further evaporation replacing the classical singularity with a high-density core governed by LQG dynamics. This core, which would be the Planck star, has an interior governed by a bounce solution that prevents divergence of curvature invariants and terminates further collapse or explosive expansion. The outer geometry still satisfies Schwarzschild asymptotics but the quantum-corrected interior acts as a stable relic, consistent with the idea that Hawking evaporation ceases once the Planck regime is reached due to breakdown of semiclassical approximations.
\\
\\
In our model, the interior of the collapsing homogeneous sphere follows the dynamics dictated by LQC, which predicts a bounce at the critical density $ \rho_c \sim \rho_{\text{Planck}} $. As shown in \S 2, the scale factor $a(t)$ never vanishes and the areal radius $r(t) = r_0 a(t) $ reaches a non-zero minimum before beginning to expand. However, for the expansion to be visible externally, matter must re-emerge from within the Schwarzschild radius. This is something that is forbidden by classical GR and ruled out by observations, unless spacetime structure itself is radically altered. So we are naturally led to interpret the bounce as giving rise not to a re-expanding visible object but rather to a Planck star, which is a compact region that undergoes a quantum bounce internally but remains causally hidden from external observers.
\\
\\
One can also ponder that an expansion would lead to an explosion in that regime because Einstein's event horizon is not a barrier to escape when the size of the remnant is of order the Schwarzscild radius. We remark here that it is indeed true that if the bounce radius equals the Schwarzschild radius, then in classical GR the event horizon disappears and matter could in principle re-expand visibly. However, in our scenario, the effective dynamics are governed by quantum geometry near Planck scales and the would-be horizon is replaced by a non classical surface where quantum effects dominate. We assume that the post bounce state does not lead to explosive dispersal of energy but instead settles into a frozen Planck-scale configuration, which would be the Planck star remnant, due to the balance between quantum pressure and residual gravitational self-attraction. This avoids re-expansion beyond the horizon, rendering the object unobservable and consistent with astrophysical constraints. The remnant remains stable and hidden, as no causal signal can escape in finite time due to infinite redshifting near the Schwarzschild radius.
\\
\\
Our analysis began by modeling the interior geometry of the collapsing matter using a homogeneous and isotropic FLRW metric. While the FLRW metric is often used to describe the entire universe in cosmological contexts, we do not assume here that the matter within the collapsing interior region must necessarily be constituting a universe unto itself. Rather we employ the FLRW ansatz as a symmetry-reduced, idealized description of the interior evolution of a localized, spherically symmetric region undergoing gravitational collapse. This choice is suitable to primordial black holes and their quantum gravitational remnants in the early universe. PBHs are expected to form from large-amplitude perturbations that re-enter the Hubble horizon during the radiation-dominated epoch \cite{pbh14green2024primordial}. These perturbations typically generate nearly spherical, homogeneous overdensities, and these in the absence of significant anisotropies or angular momentum evolve in a manner that is effectively captured by a closed FLRW metric.
\\
\\
The quantum bounce mechanism provided by LQC then replaces the classical singularity with a finite-density core, leading to the formation of PSR. This approach offers a proper means of modeling the internal dynamics of such remnants and forms a key point of our argument towards them being possible dark matter candidates.
\\
\\
The collapse halts at radius $ R \sim \ell_{\text{Planck}} $ and mass $ M \sim M_{\text{Planck}} $, forming a non-singular core that no longer evolves on observable timescales. This core does not radiate, does not collapse further, and does not explode, and so it simply remains as a stable relic. To see this quantitatively, we note that the maximum energy density in the bounce scenario is $ \rho_c \sim \rho_{\text{Planck}} = \frac{c^5}{h G^2} \sim 10^{93}  \text{g~cm}^{-3} $. The mass contained within a volume of Planck length $ \ell_{\text{Pl}}^3 $ at this density is
$ M_{\text{Pl}} = \sqrt{h c / G} \sim  10^{-5}  \text{g} $. This yields a compact object of size $ \sim 10^{-33}  \text{cm} $ and mass $ \sim 10^{-5}  \text{g} $, with Schwarzschild radius matching its physical radius:
\begin{equation}
    R_s = \frac{2GM}{c^2} \sim \ell_{\text{Pl}}.
\end{equation}
Thus, the Planck star is naturally stabilized by quantum effects at the Planck scale.
\\
\\
Such relics could arise from PBHs in the early universe as PBHs with initial masses $ M_0 \lesssim 10^{15}  \text{g} $ would have evaporated by now via Hawking radiation, potentially leaving behind PSR. If such remnants were produced in large numbers in the early universe, they would now constitute a cold, non-luminous, and effectively collisionless dark matter component and given this, we can estimate the required relic abundance. The present-day dark matter density is
\begin{equation}
    \rho_{\text{DM}} \approx 1.3 \times 10^{-6}  \text{GeV/cm}^3 \approx 2.3 \times 10^{-30}  \text{g~cm}^{-3} .
\end{equation}
Assuming that Planck star relics have a mass $ M_{\text{Pl}} \sim  10^{-5} \text{g} $, the number density required to match dark matter is
\begin{equation} \label{nden}
    n_{\text{relic}} = \frac{\rho_{\text{DM}}}{M_{\text{Pl}}} \sim \frac{2.3 \times 10^{-30}}{10^{-5}} \sim 2.3 \times 10^{-25}  \text{cm}^{-3}.
\end{equation}
This is a very small number density, implying only about 200 relics within the entire volume of the Earth. Such a population would evade all direct detection constraints, yet could gravitationally contribute to galactic and cosmic structures.
\\
\\
To further assess the viability of Planck star remnants as dark matter candidates, we consider additional astrophysical and cosmological quantities. For instance, let us estimate the total number of Planck relics required to account for the observed dark matter density in the universe. Assuming a comoving volume of the observable universe
\begin{equation}
V_{\text{obs}} \sim \frac{4}{3} \pi R_{\text{H}}^3 \sim \frac{4}{3} \pi (4.4 \times 10^{28}  \text{cm})^3 \sim 3.6 \times  10^{86}  \text{cm}^3,
\end{equation}
the total number of PSR relics required using \eqref{nden} is
\begin{equation}
N_{\text{relic}} = n_{\text{relic}} \times V_{\text{obs}} \sim (2.3 \times  10^{-25} ) \times (3.6 \times  10^{86}) \sim 8.3 \times  10^{61}.
\end{equation}
Interestingly, this number is 27 orders of magnitude below the estimated number of particles (e.g. photons) in the universe ($\sim 10^{88}$) which would make the relic population extremely sparse.
\\
\\
In considering the cosmological viability of PSR as dark matter candidates, it is important to also check whether such objects overcontribute to the entropy content in the universe. Although classical black holes radiate thermally via Hawking radiation with a characteristic temperature
\begin{equation}
T_{\text{H}} = \frac{1}{8\pi G M},
\end{equation}
this expression becomes increasingly uncertain as $M \to M_{\text{Pl}}$, where quantum gravity effects dominate and semiclassical approximations break down. In the Planck star scenario, the evaporation process does not proceed indefinitely \cite{pl1Rovelli:2014cta,pl2Rovelli:2024sjl} and instead, it halts when the black hole reaches a critical density or curvature threshold leading to a remnant of Planck mass. While such a remnant is no longer expected to emit Hawking radiation, for cosmological considerations we can still treat the final Hawking temperature prior to stabilization as an upper bound on the effective energy scale associated with the remnant at the time of its formation. Assuming that evaporation ends at a remnant mass $M_{\text{Pl}}$, the maximum temperature attained by the object during its evolution is roughly
\begin{equation}
T_{\text{freeze}} \sim \frac{1}{8\pi G M_{\text{Pl}}} \sim 10^{32} \text{K}.
\end{equation}
This temperature characterizes the energy scale of the remnant at the end of evaporation. As the universe expands, the kinetic energy of non-relativistic remnants like PSR redshifts as \( 1/a^2 \) and so even if the relic had a thermal history, its effective "temperature" today would be suppressed accordingly which would be leading to
\begin{equation} \label{temp}
T_{\text{today}} \sim T_{\text{freeze}} \left( \frac{a_{\text{evap}}}{a_0} \right)^2 .
\end{equation}
For relics formed in the very early universe, the dilution factor would be enormous. It would easily be exceeding $10^{64}$, which implies that the relics today would constitute cold dark matter. To better justify the cold nature of PSR we note that while these relics are non-radiating and non-interacting after formation and their "coldness" in a cosmological sense refers specifically to their negligible velocity dispersion and short free-streaming length. Any residual thermal energy from their evaporation era formation would be enormously redshifted by today with the suppression factor in \eqref{temp}, making their kinetic contribution to structure formation negligible and so this places them well within the cold dark matter regime. Moreover as the relic does not radiate or interact significantly after freeze-out, it behaves as pressureless matter and this ensures compatibility with cosmological observations. It is particularly in line with those constraining the cosmic microwave background and effective number of relativistic species $N_{\text{eff}}$ \cite{Planck:2018vyg}. The entropy content of such relics is also likewise negligible compared to the thermal entropy of the cosmic microwave background or the cosmic neutrino background.
\\
\\
\begin{figure}[!h]
    \centering
    \includegraphics[width=1\linewidth]{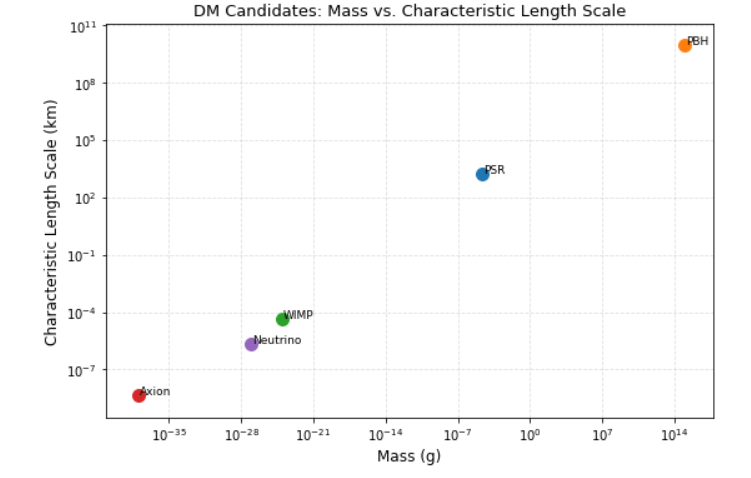}
    \caption{Mass versus characteristic separation of dark matter candidates, including Planck Star Remnants (PSR).}
    \label{dml}
\end{figure}
In figure \ref{dml} we present a comparative plot of dark matter candidates in the mass–length scale plane. It is interesting to note that PSR occupy an intermediate region of this parameter space in the sense that they are significantly more massive than WIMPs or axions \footnote{In this figure we have plotted a representative axion mass of order $\sim 10^{-5}$eV which is consistent with conventional QCD axion models but we do acknowledge that axion-like particles in string inspired scenarios or other extensions could span a vastly broader mass range, from $10^{-22}$eV up to the MeV scale \cite{adm1Ringwald:2024uds,adm2Choi:2020rgn,adm3Chadha-Day:2021szb}.} yet considerably lighter than primordial black holes, and their spatial distribution is neither highly clustered nor extremely rarefied. This balance positions PSR in a compelling "sweet spot" with respect to mass and spatial abundance, which could potentially be allowing them to fulfill dark matter constraints without conflicting with structure formation or observational bounds.
\\
\\
We note that Planck mass relics have been discussed previously as potential dark matter candidates, most notably by MacGibbon \cite{mac1MacGibbon:1987my} and follow-up  work \cite{mac2Maldacena:2020skw,mac3Rasanen:2018fom,mac4Barrau:2019cuo,mac5Chen:2004ft,mac6XENON:2023iku,ref1Arbey:2021mbl,ref2Easson:2002tg,ref3Pacheco:2018mvs,ref4Calza:2024fzo,ref5Calza:2024xdh,ref6Dialektopoulos:2025mfz} and the consideration of remnant type dark matter was recently seen in \cite{Davies:2024ysj}. The previous models state that Hawking evaporation ceases  once the mass reaches the Planck scale, leaving behind stable Planck mass relics. The motivation for such models is largely thermodynamic or based on semi-classical arguments for UV cutoffs without much said about the structure of these relics. In contrast, the PSR discussed in this work are compact objects with Schwarzschild-scale geometry and are derived from an explicit quantum gravitational scenario in LQG. The PSR arises when the classical singularity in gravitational collapse is resolved due to quantum geometry effects, resulting here in a bounce at the Planck density. The internal spacetime in our case is governed by an effective LQC-modified FLRW geometry, and the exterior is matched to a Schwarzschild metric. This is a fundamentally different picture than that envisioned by MacGibbon-type relics, which do not incorporate bounce dynamics or any non-trivial internal geometry.
\\
\\
There are important quantitative and qualitative differences between our PSR proposal and the MacGibbon-type relics. First, the PSR in our model have a Schwarzschild radius corresponding to their mass, 
\begin{equation*}
    R_{\text{PSR}} = \frac{2GM_{\text{Pl}}}{c^2} \sim  10^{-33}~\text{cm} ,
\end{equation*}
and contain an LQC-governed FLRW interior with a bounce that replaces the singularity. This internal geometry is physically meaningful with the rich quantum gravity effects and provides a richer dynamical history for the remnant.Secondly, in the MacGibbon type relics one must impose an external constraint to ensure the stability of relics, either through new symmetries or assumptions that quantum gravity provides an energy barrier to decay. PSR  are naturally stable due to their origin in Loop quantum effects, where the bounce leads to a halting of collapse at finite density and the object remains gravitationally bound, causally disconnected from external observers, and thus shielded from decay channels. Furthermore, they do not require any exotic conservation laws or speculative symmetries.
\\
\\
\\
\\
We also note that bouncing solutions have been considered in other recent contexts, such as in the work by Gaztañaga et al.\cite{bounceexclusion} which introduces a bounce induced by a quantum exclusion principle acting as a relativistic degeneracy pressure. While that model is focused on avoiding singularities in gravitational collapse, it differs fundamentally from our approach in several respects as firstly the bounce mechanism in their framework is not derived from a full quantum gravity theory but is postulated phenomenologically, whereas our model is rooted in the effective dynamics of loop quantum effect. Second, their solution leads to an inflationary expansion that fills a cosmological horizon-sized volume, effectively suggesting a universe-scale outcome of collapse while in contrast our scenario terminates in a stable PSR. Finally, while their work aims to account for cosmological observables such as spatial curvature and the CMB quadrupole anomaly, our analysis emphasizes astrophysical consequences including relic abundances, structure formation bounds etc. relevant to the dark matter sector.
\section{Conclusions}

PSR behave as cold dark matter contributing to the formation of large-scale structures. Their lack of electromagnetic interaction and low number density could make them ideal candidates for constituting the bulk of dark matter. From the standpoint of particle phenomenology they differ from WIMPs, axions, or sterile neutrinos and from the gravitational side they seemingly satisfy all necessary criteria. Our work strengthens this picture by showing that quantum-regulated collapse leads generically to such remnants under physically reasonable assumptions. Once the bounce occurs, a relic would remain. The relic would be stable, non-radiating, and non-interacting, and it carries the gravitational imprint of its formation epoch. 
\\
\\
The primary emphasis of earlier proposals of Planck Stars was on conceptual resolution of the singularity and information loss problems \cite{pl1Rovelli:2014cta,pl2Rovelli:2024sjl}. These works also considered white hole remnants formed via black hole evaporation and even suggested that Planck-scale cores might tunnel into expanding white holes over long timescales \cite{pl3Rovelli:2017zoa,pl4Scardigli:2022jtt}. Our approach is substantially different as we implement tools from loop quantum theory directly into the collapsing interior geometry and consider its matching to a Schwarzschild exterior via Israel junction conditions. This framework allows us to demonstrate that a bounce occurs at Planckian density, halting collapse and there could be a formation of a non-singular PSR in this case. Moreover while others have considered a range of small masses \cite{pl5Wilson-Ewing:2024uad,pl6Barrau:2014hda,pl7Tarrant:2019tgv,Planck:2018vyg}, our model considers a fixed Planck-scale mass for the PSR and this  places the remnant in a well-defined region of parameter space with calculable number density and gravitational signatures. A viable question which may also arise could be about the observational features of PSR and whether one can expect such features to even exist. In this regard, we would like to remark that a follow up work of ours on this subject addresses this aspect aptly as well. 
\\
\\
We would also like to note that in our scenario the key idea is that PSR is not treated as a generic or structureless endpoint, but as the result of a well defined dynamical bounce described by effective Loop quantum dynamics. The internal structure of the remnant encodes a non trivial geometric history and to be specific, it lies in the bounce evolution of the collapsing matter. In this sense the PSR is not a degenerate endpoint of many initial conditions but a unique outcome of a unitary evolution governed by quantum gravitational corrections. The entropy that accumulates in the Hawking radiation reflects entanglement with the interior degrees of freedom of the remnant, which are governed by a high-density quantum geometry \footnote{For some details of the entropy structure for Hawking radiation, see \cite{Almheiri:2020cfm,Mathur:2009hf}.}. While we do not yet have a complete microscopic accounting of the remnant’s entropy, the LQC framework in general implies that the bounce retains information about the initial state and this opens up the possibility that the fine-grained entropy of the remnant need not be small even though its geometric entropy (in the Bekenstein sense) is. Our picture would thus be consistent with a pure global state when both radiation and interior remnant are considered together. It may even be an interesting endeavor to pursue a different work which is focused on the details of the entropy structure of PSR, particularly treating its entropy in LQG consistent ways.
\\
\\
\section*{Acknowledgements}
OT was supported in part by the Vanderbilt Discovery Alliance Fellowship. AL was supported in part by the Black Hole Initiative, which is funded by GBMF anf JTF.

\bibliography{references}
\bibliographystyle{unsrt}

\end{document}